\documentclass[aps,pre,floats,superscriptaddress,showpacs,twocolumn]{revtex4}

\usepackage{amsmath, amsthm, amssymb}
\usepackage{enumitem}
\usepackage{graphics,graphicx}
\usepackage{epsfig}
\usepackage{times}
\usepackage{dcolumn}
\usepackage{bm}

\usepackage{ulem, color}

\begin{document}

\title{Enhancing the stability of the synchronization of multivariable coupled oscillators}

\author{R. Sevilla-Escoboza}
\affiliation{Centro Universitario de los Lagos, Universidad de Guadalajara, Enrique D\'{i}az de Leon, Paseos de la Monta\~na, Lagos de Moreno, Jalisco 47460, Mexico}
\author{R. Guti\'errez}
\affiliation{Department of Chemical Physics, Weizmann Institute of Science, Rehovot 76100, Israel}
\author{G. Huerta-Cuellar}
\affiliation{Centro Universitario de los Lagos, Universidad de Guadalajara, Enrique D\'{i}az de Leon, Paseos de la Monta\~na, Lagos de Moreno, Jalisco 47460, Mexico}
\author{S. Boccaletti}
\affiliation{CNR-Istituto dei Sistemi Complessi, Via Madonna del Piano, 10, 50019 Sesto Fiorentino, Italy}
\affiliation{The Italian Embassy in Israel, 25 Hamered Street, 68125 Tel Aviv, Israel}
\author{J. G\'omez-Garde\~nes}
\affiliation{Departamento de F\'{\i}sica de la Materia Condensada, University of Zaragoza, Zaragoza 50009, Spain}
\affiliation{Institute for Biocomputation and Physics of Complex Systems (BIFI), University of Zaragoza, Zaragoza 50018, Spain}
\author{A. Arenas}
\affiliation{Departament d'Enginyeria Inform\`atica i Matem\`atiques, Universitat Rovira i Virgili, 43007 Tarragona, Spain}
\author{J.M. Buld\'u}
\affiliation{Laboratory of Biological Networks, Center for Biomedical Technology, UPM, Pozuelo de Alarc\'{o}n, 28223 Madrid, Spain}
\affiliation{Complex Systems Group, Universidad Rey Juan Carlos, 28933 M\'ostoles, Madrid, Spain}

\date{\today}

\begin{abstract}
Synchronization processes in populations of identical networked oscillators are in the focus of intense studies in physical, biological, technological and social systems. Here we analyze the stability of the synchronization of a network of oscillators coupled through different variables. Under the assumption of an equal topology of connections for all variables, the master stability function formalism allows assessing and quantifying the stability properties of the synchronization manifold when the coupling is transferred from one variable to another.  We report on the existence of an optimal coupling transference that maximizes the stability of the synchronous state in a network of R\"ossler-like oscillators. Finally, we design an experimental implementation (using nonlinear electronic circuits) which grounds the robustness of the theoretical predictions against parameter mismatches, as well as against intrinsic noise of the system.
\end{abstract}

\maketitle

\section{Introduction}

Synchronization processes on complex networks has received a lot of attention during the last decades \cite{boccaletti2006,arenas2008,barrat2008}. The interplay between the dynamical evolution of oscillators and their local interactions (as given by the complex topology of a network) usually results in non-trivial phenomena of relevance to physical, biological, technological and social systems. First introduced by Pecora and Carroll \cite{pecora1998}, the Master Stability Function (MSF) is nowadays one of the main theoretical methods for the study of network synchronization. MSF is indeed a powerful tool to analyze the stability of the synchronization manifold when identical systems of oscillators are diffusively coupled.
Originally applied to undirected networks, the MSF approach has been later extended to investigate enhancements and optimization of complete synchronization in weighted and asymmetric topologies (see \cite{boccaletti2006,arenas2008}, and references therein).

In \cite{motter2005a} the authors stated the so-called heterogeneity paradox, i.e. the fact that heterogeneous networks, wherein distances between nodes are relatively short, are less stable, in terms of synchronization, than their homogeneous counterparts. Soon after, a proper and adequate weighting of the link strengths was shown to overcome this paradox, based again on concepts sparkling from the MSF formalism \cite{chavez2005, motter2005b}. Following works, have shown how different network's  topological features influence the stability of the synchronous state, such as: heterogeneity of the node degree, degree-degree correlations, average shortest-path, betweenness centrality or clustering. These latter studies indicate that altering the structure of a network may result in maximizing  the stability of the synchronous state, thus achieving a {\it maximally stable synchronization structure} \cite{nishikawa2005}. Enhancement of the networks' synchronizability can also be achieved by the application of genetic algorithms increasing the stability of the synchronized state. In this case, the networks self-organize by disconnecting the hubs and connecting peripheral nodes, thus increasing the homogeneity  and leading to what is known as {\it entangled networks} \cite{donetti2005}.

\begin{figure}
\begin{center}
\includegraphics[width=0.34\textwidth, angle=-90]{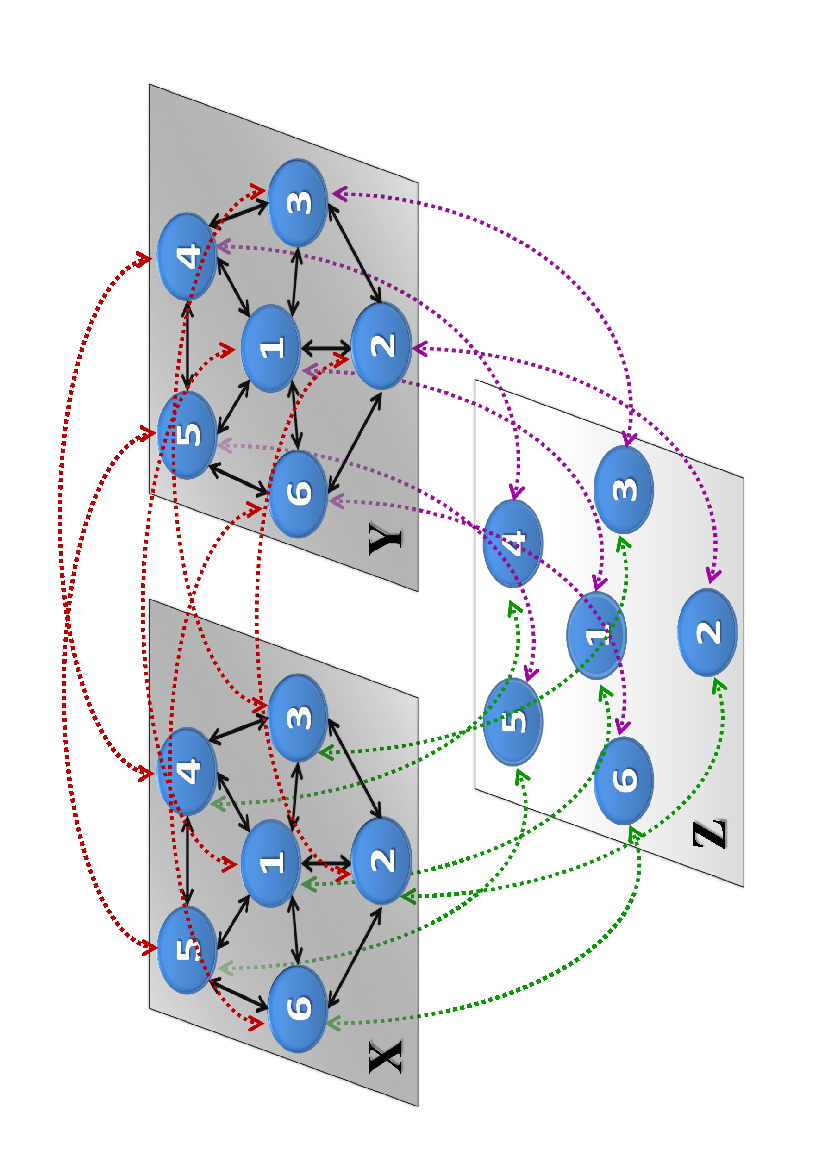}
\end{center}
\vspace{-1.0cm}
\caption{\label{fig:f03} (Color
 online) Network of $N=6$ dynamical systems coupled through different variables.
In order to better observe the coupling introduced at each variable, each node $i$ is split into different layers, each one corresponding to a variable of the system ($X$, $Y$ and $Z$), whose complete dynamical state is given by the vector $\mathbf{x}=(x,y,z)$ obtained from the combination of the states of its variables at each layer. Note that, in this particular example, only variables $x$ and $y$ are used to couple the systems and, in turn, that the topology of the coupling network is the same for both variables.
}
\label{fig01}
\end{figure}

In our study, we report the enhancement of the stability of complete synchronization of an ensemble of dynamical units, when coupled simultaneously in different dimensions.
We are concerned with a multivariable coupling, where the dynamical systems are coupled through different dimensions according to
a certain structure of connections (see Fig.\ref{fig01} for a schematic illustration).
In particular, we consider a generic dynamical system whose associated vector state $\mathbf{x}$ (with  $\mathbf{x} \in \mathbb{R}^m$) evolves according to $\dot{\bf x} = {\bf f}({\bf x})$.
Each one of the $m$ state variables of the dynamical system at a given node can be coupled to the corresponding variable of any of the other
systems (i.e., nodes) of the network. 

Equivalently, we can think of our system as a network with $l\leq m$ layers, each one accounting for the structure of couplings at each variable of the system. This multilayer point of view illustrated in  Fig.\ref{fig01} is, in fact, just accounting for a multivariable coupling between the nodes of a network, nevertheless it will help us to provide a more concrete representation of the structure of the system, and possible connections to applications. So we will make use of it at certain points. If the coupling between oscillators does not include some of the state variables, {\it i.e.} $l < m$, the topology of the corresponding layers to those variables would be trivially given by a zero adjacency matrix, so we would not consider them to be proper layers (as is the case of the layer corresponding to variable $z$ in Fig.\ref{fig01}). For simplicity, we consider a bidirectional coupling between the same variables of each system (i.e. each layer is an undirected network). This is illustrated in Fig. \ref{fig01} with an example of the case $l=2$ and $m=3$.


Interestingly, our framework connects with the so-called {\it hypernetwork} formalism introduced by Sorrentino \cite{sorrentino2012}. In this latter work, the author shows that a MSF approach to hypernetwork synchronization is possible when the Laplacian matrices
\cite{laplacian} of different layers (accounting for the coupling through each variable)  have the same basis of eigenvectors, {\it i.e.}, when they are simultaneously diagonalizable. This is a condition that has been shown to be fulfilled for two layers in three cases: (i) the Laplacian matrices of the different layers are commuting (a condition that automatically allows for a MSF approach whatever the number of layers if the Laplacian matrices form a pairwise commuting set), (ii) one of the two layers is unweighted and fully connected, or (iii) one of the two layers has an adjacency matrix of the form $A_{ij}=a_j$ with $i,j=1,...,N$. Additionally, Ref. \cite{irving2012} contains an extension of the approach in \cite{sorrentino2012} to more general topologies by making use of a simultaneous block diagonalization of Laplacian matrices corresponding to different layers, thereby decreasing the dimensionality of the linear stability problem.

In our work, we consider the topology to be the same in each layer, trivially falling, from the of view of hypernetworks, into category (i) of Ref. \cite{sorrentino2012}. This way, we present a study on how the stability of the synchronous state is enhanced by finding an optimal balance for the coupling between the different variables in a network of identical oscillators with multivariable coupling. On the one hand, we provide results based on extensive numerical simulations of networks of R\"ossler-like oscillators to show the applicability of the proposed ideas and how the MSF can help us to find the adequate balance between the couplings that optimizes the stability of the synchronous state of a network. On the other hand, by constructing an electronic version of the model, we show that these predictions are in good agreement with the experimental evidences in spite of the idealizations used in the theoretical treatment.

\section{Coupling through different variables with identical topology}
\label{sec:MSF}

In this section we explain how stability of the synchronous state can be enhanced by engineering a multivariable coupling function between nodes in a network and what balance between coupling variables is the most adequate. For the sake of concreteness we focus on a set of R\"ossler-like oscillators \cite{modelo} coupled to their neighbors through both the $x$ and the $y$ variables, whose dynamics evolve according to the following equations:
\begin{eqnarray}\label{system}\nonumber
\dot{x}_i&=&-\alpha_{1}(x_i+\beta y_i+\Gamma z_i-(1-d_r)\sigma_X\psi \sum_{j=1}^N{a}^X_{ij} \left[ x_j-x_i%
\right] )  \\  \nonumber
\label{system1}
\dot{y}_i&=&-\alpha_{2}( -\gamma x_i+\left( 1-\delta \right)
y_i -d_r \sigma_Y\phi \sum_{j=1}^Na^Y_{ij} \left[ y_j-y_i\right])  \\
\label{system2}
\dot{z}_i&=&-\alpha_{3}\left( G_{x_i}+z_i\right)\;,
\end{eqnarray}
where $\alpha_{1}=500$, $\alpha_{2}=200$, $\alpha_{3}=10,000$, $\beta =10$, $\Gamma =20$, $\gamma =50$, $\delta =8.772$, $\mu=15$, $\psi =20$, $\phi =50$ and
$(1-d_r) \sigma_X$ and $d_r \sigma_Y$ account for the coupling strengths of variables $x$ and $y$. As we explain below, this chaotic oscillator has the highly non-trivial characteristic of being quite robust when implemented in electronic circuits. The adjacency matrices $\mathcal{A}^X$ and $\mathcal{A}^Y$ contain the topology of each of two layers, each one accounting for the coupling through the $x$ and $x$ variables. Elements $a^X_{ij}$ and $a^Y_{ij}$ are one when nodes $i$ and $j$ are connected and zero otherwise.
With these parameters the oscillators display chaotic dynamics due to the nonlinearity introduced in $G_{x_i}$, which consists on  a piecewise function defined as:
\begin{eqnarray}
G_{x_i} &=&\left\{
\begin{array}{cc}
0 & x_i\leq 3 \\
\mu \left( x_i-3\right)  & x_i>3%
\end{array}%
\right.\;
\end{eqnarray}

The coupling between oscillators is here controlled by two parameters: $\sigma$ being the coupling strength and $d_r$ controlling how the coupling strength is distributed between variables $x$ and $y$. This way, $d_r=0$ ($d_r=1$) leads to a coupling restricted to variable $x$ ($y$), while a sweep of $d_r$ in the interval $[0,1]$ allows for a weighted combination of both $x$ and $y$ variables. Notice that the role of the parameter $d_r$ is therefore that of exploring the consequences of unevenly distributed coupling on the stability of synchronization, which in the past were the object of specific studies in space extended systems \cite{bragard2003} and weighted monolayer graphs \cite{hwang2005}. Here however, the difference in the weight assigned to each variable introduced by $d_r$ implies a different balance of two system's variables in the coupling configuration, but does not affect the un-directionality of each one of the network's links.

We now apply the MSF formalism to study how the amount of coupling  effectively distributed among the two coupling variables affects the stability of the synchronous state of the network.
Denoting the coupling functions of each variable as $\textbf{h}_{X,Y} : \mathbb{R}^m \rightarrow \mathbb{R}^m$ \textcolor{blue}{, t}he dynamics of each node is then  given by:

\begin{widetext}
\begin{eqnarray}\label{msfsystem}
\nonumber
\dot{\mathbf{x}}_i &=& \textbf{f}(\mathbf{x}_i)\ +\ (1-d_r) \sigma_X \displaystyle\sum_{j=1}^N \mathcal{A}^X_{ij}\left[\, \textbf{h}_X(\mathbf{x}_j) -\textbf{h}_X(\mathbf{x}_i) \right] +\ d_r \sigma_Y \displaystyle\sum_{j=1}^N \mathcal{A}^Y_{ij}\left[\, \textbf{h}_Y(\mathbf{x}_j) -\textbf{h}_Y(\mathbf{x}_i) \right]\\
\nonumber
&=& \textbf{f}(\mathbf{x}_i)\ +\ (1-d_r) \sigma_X \displaystyle\sum_{j=1}^N \left(\mathcal{A}^X_{ij} - \delta_{ij} \left[\displaystyle\sum_{j=1}^N\mathcal{A}^X_{ij}\right]\right) \textbf{h}_X(\mathbf{x}_j) +\ d_r \sigma_Y \displaystyle\sum_{j=1}^N \left(\mathcal{A}^Y_{ij} - \delta_{ij} \left[\displaystyle\sum_{j=1}^N\mathcal{A}^Y_{ij}\right]\right) \textbf{h}_Y(\mathbf{x}_j)\\
\nonumber
&=& \textbf{f}(\mathbf{x}_i)\ -\ (1-d_r) \sigma_X \displaystyle\sum_{j=1}^N \mathcal{L}^X_{ij} \textbf{h}_X(\mathbf{x}_j)  -\ d_r \sigma_Y \displaystyle\sum_{j=1}^N \mathcal{L}^Y_{ij} \textbf{h}_Y(\mathbf{x}_j)\\
\label{fullsys}
\end{eqnarray}
\end{widetext}
where $\delta_{ij}$ stands for the Kronecker delta, and $\mathcal{L}^{X,Y}_{ij}$ are the Laplacian matrices \cite{laplacian} describing the coupling through
variables $x$ and $y$ respectively.
If we consider that $\sigma_Y=\sigma_X=\sigma$ and we restrict our analysis to the case of identical coupling topologies for all variables of the system,
i.e. $\mathcal{A}^X=\mathcal{A}^Y=\mathcal{A}$, and, in turn, $\mathcal{L}^X=\mathcal{L}^Y=\mathcal{L}$, Eq. \ref{fullsys} reads:
\begin{equation}
\dot{\mathbf{x}}_i = \textbf{f}(\mathbf{x}_i)\ -\ \sigma \displaystyle\sum_{j=1}^N \mathcal{L}_{ij} \textbf{h}(\mathbf{x}_j)\\
\label{syssimpl}
\end{equation}
where the coupling function is $\textbf{h}(\mathbf{x}_j)= (1-d_r) \textbf{h}_X(\mathbf{x}_j)- d_r \textbf{h}_Y(\mathbf{x}_j)$. This way, Eq. \ref{syssimpl} is basically the
classical equation describing the evolution of a diffusively coupled systems, with the particularity that the coupling function $\textbf{h}$ depends on the parameter
$d_r$ accounting for how the total amount of coupling is divided between the coupling variables of the system.
The dependence on $d_r$ leads to a {\it parametric} MSF that describes the stability of the synchronization manifold. Varying the value of $d_r$ we obtain a family of MSFs that allows to evaluate how the stability of the synchronized manifold is affected by shifting the coupling from one layer to the other.

The independent variable $\nu$ of the MSF is related with the eigenvalues of the Laplacian matrix  ($\nu \equiv \sigma \gamma_k$) and the synchronization manifold will be stable when, for all eigenvalues of the Laplacian matrix $\gamma_k$, the corresponding $\nu$ leads to a value of the MSF that is negative (for a given value of $\sigma$). Taking into account that $\gamma_1=0$ (since the Laplacian is a zero row sum matrix) and that $\gamma_2 \leq \gamma_3 \leq ... \leq \gamma_N$, dynamical systems can be classified, depending on the shape of its corresponding MSF. The classification includes: {\it i)} class I systems, whose MSF is always positive for any value of $\nu$ so that the system cannot be synchronized for any topology or coupling strength, {\it ii)} class II systems, when the MSF is positive for low values of $\nu$ and becomes negative when a threshold value $\nu_I$ is achieved (being the synchronization manifold stable when $\sigma \gamma^X_2>\nu_I$) and {\it iii)} class III, when the MSF has two zeros at $\nu_I$ and $\nu_{II}$, leading to a stable synchronized manifold when $\nu_I<\sigma \gamma_2$ and $\nu_{II}> \sigma \gamma_N$ simultaneously.

\begin{figure}
\begin{center}
\includegraphics[scale=0.295,angle=90]{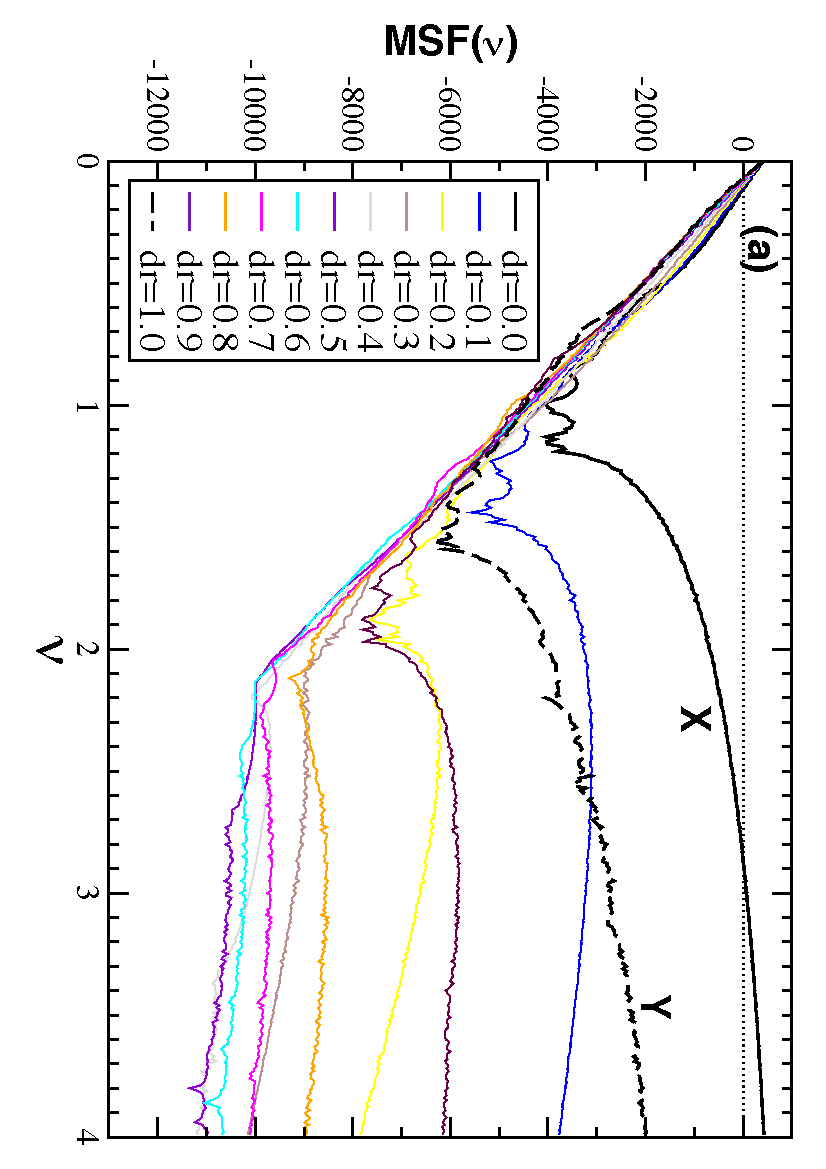}
\includegraphics[scale=0.295,angle=90]{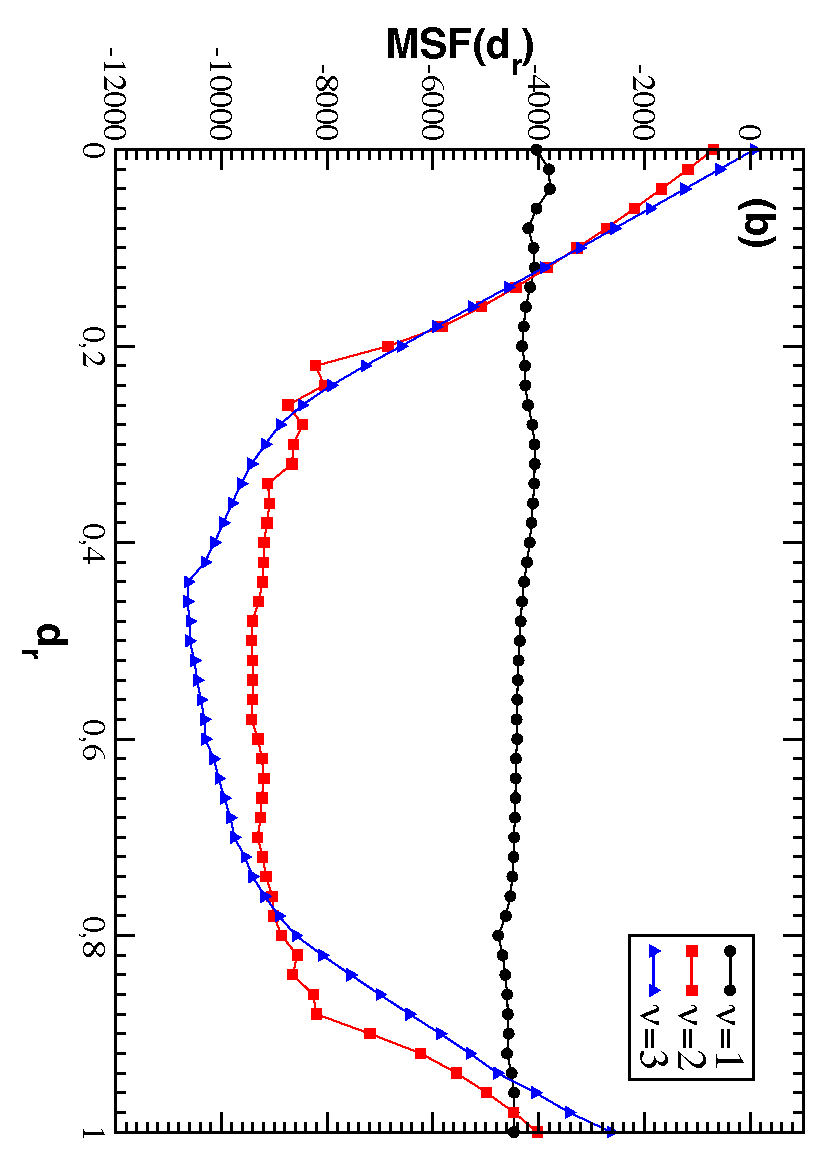}
\end{center}
\caption{\label{fig02} (Color
 online) Master Stability Function MSF($\nu$) as a function of the coupling fraction $d_r$ between the variables to be coupled. In (a), solid and dashed black lines are, respectively, the MSF for a coupling through variables $x$ and $y$ exclusively. Colored lines show the MSF obtained for different values of $d_r$, as indicated in the legend. In (b), MSFs for three different values of the parameter $\nu$:
$\nu=1$, $\nu=2$ and $\nu=3$. Note that, for a sizable range of $d_r$, the multivariable coupling leads to lower values of the MSF, indicating a region where the stability against external perturbations is higher.}
\end{figure}

Now we investigate the synchronization properties of the system depending on the distribution of the coupling strengths among the coupling variables. In Fig. \ref{fig02}(a) we show the MSF obtained for the R\"ossler-like system coupled through the  $x$ variable ($d_r=0$, black solid line), the $y$ variable ($d_r=1.0$, black dashed line) and simultaneously the $x$ and $y$ variables (colored lines). A range of $d_r$ values is swept in order to show the gradual changes in the stability of the synchronous state as the coupling of one or the other variable is enhanced.
As  expected, coupling introduced only through variable $x$ leads the system to be class III while it becomes class II when coupled only through the $y$ variable \cite{huang2009,aguirre2014}.
Interestingly, when the coupling is introduced simultaneously through the two variables we obtain a MSF that is not a linear combination of the isolated layers. In particular, the sweep of $d_r$ leads to a family of MSFs that are class II (at least for the ten cases plotted in Fig. \ref{fig02}(a)), thus synchronizing the network when the condition $\sigma \gamma^X_2 > \nu_I$ is fulfilled, which is in principle achievable for any topology by just using a sufficiently high $\sigma$.

Figure  \ref{fig02}(a) also points out an important consequence, namely that using a multivariable coupling (while maintaining the overall coupling strength) leads to values of the MSF that are more negative than those obtained when coupling the systems through a unique variable, indicating that the stability of the synchronized
manifold is higher for multivariable couplings. This is somehow to be expected, since a simple change of coordinates would lead the dynamics of
the R\"ossler system to be described as a combination of the actual $x$, $y$ and $z$ variables. However, by studying how the MSF modifies its shape as a function
of the combination of the variables of the system we can find the regions with the highest stability.
This fact can be observed in Fig. \ref{fig02}(b), where the MSF is plotted as a function of $d_r$ for three different values of $\nu$ . We can observe a minimum of the MSF at intermediate values of $d_r$ for $\nu=2$ and $\nu=3$, {\it i.e.}, a $d_r$ value where the stability of the synchronized manifold is the largest. This result is also observed within a range of values of $\nu$ (not shown here).

\begin{figure}
\begin{center}
\includegraphics[width=0.44\textwidth]{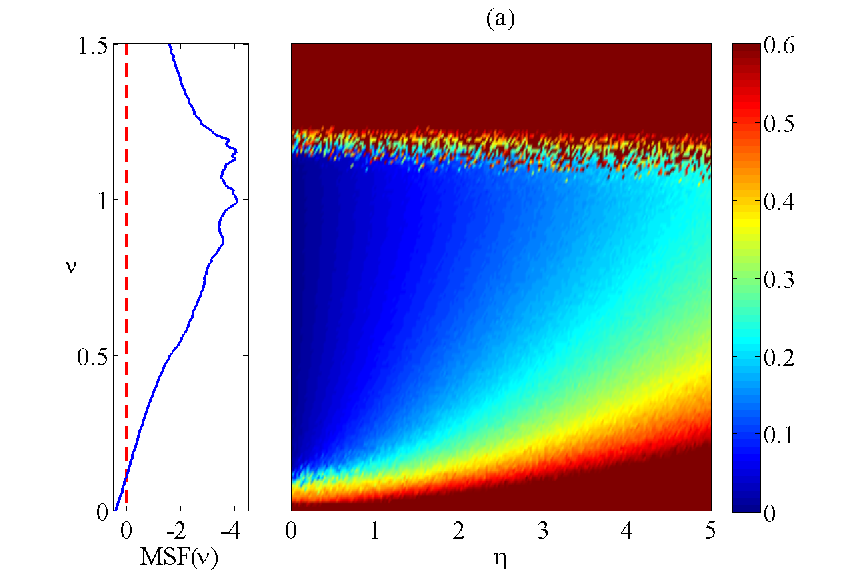}
\includegraphics[width=0.44\textwidth]{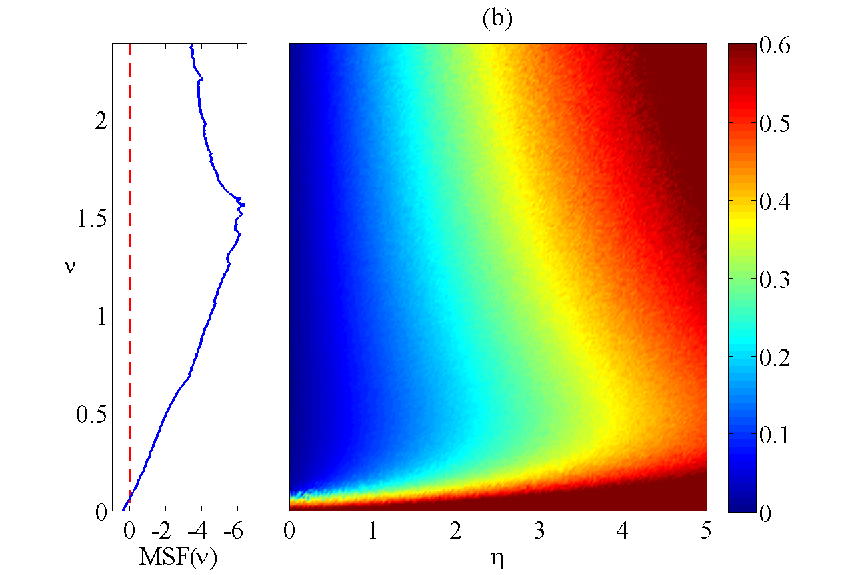}
\includegraphics[width=0.44\textwidth]{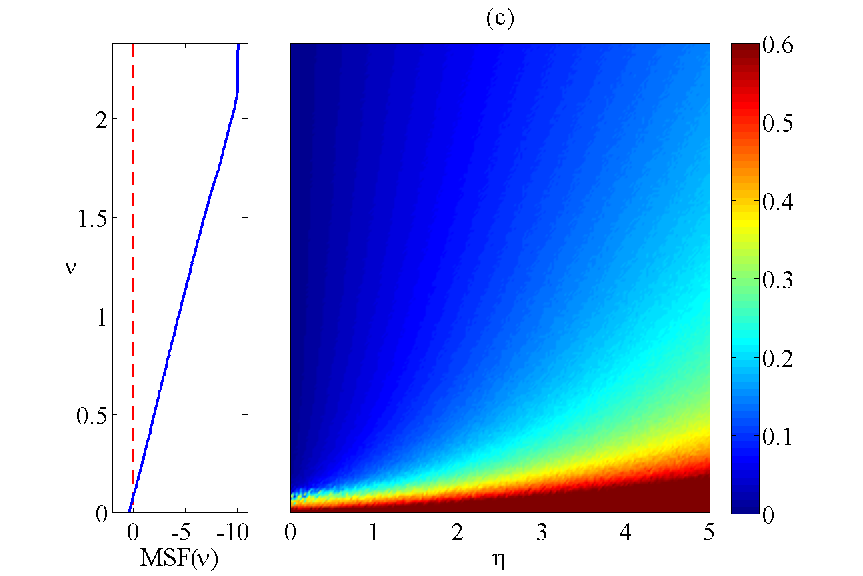}
\end{center}
\caption{\label{fig03} (Color
 online) {Robustness of the synchronized state.}
Synchronization error $\epsilon$ as a function of the noise strength $\eta$ and the coupling (rescaled as $\nu=\sigma  \gamma_2$). Panels correspond to: (a) $d_r=0$,  (b) $d_r=1$ and (c) $d_r=0.5$. The network coupled through both the $x$ and $y$ variables ($d_r=0.5$) is the one showing better performance against noise perturbations. The synchronization error in the system is computed as $\frac{2}{N\cdot (N-1)}\sum_{i<j} |x_i - x_j|$, where the normalizing factor corresponds to the total number of oscillator pairs in the network. Left plots correspond to the MSF obtained for each value of $d_r$.}
\end{figure}


\begin{figure*}[t!]
\includegraphics[scale=0.60,angle=-90]{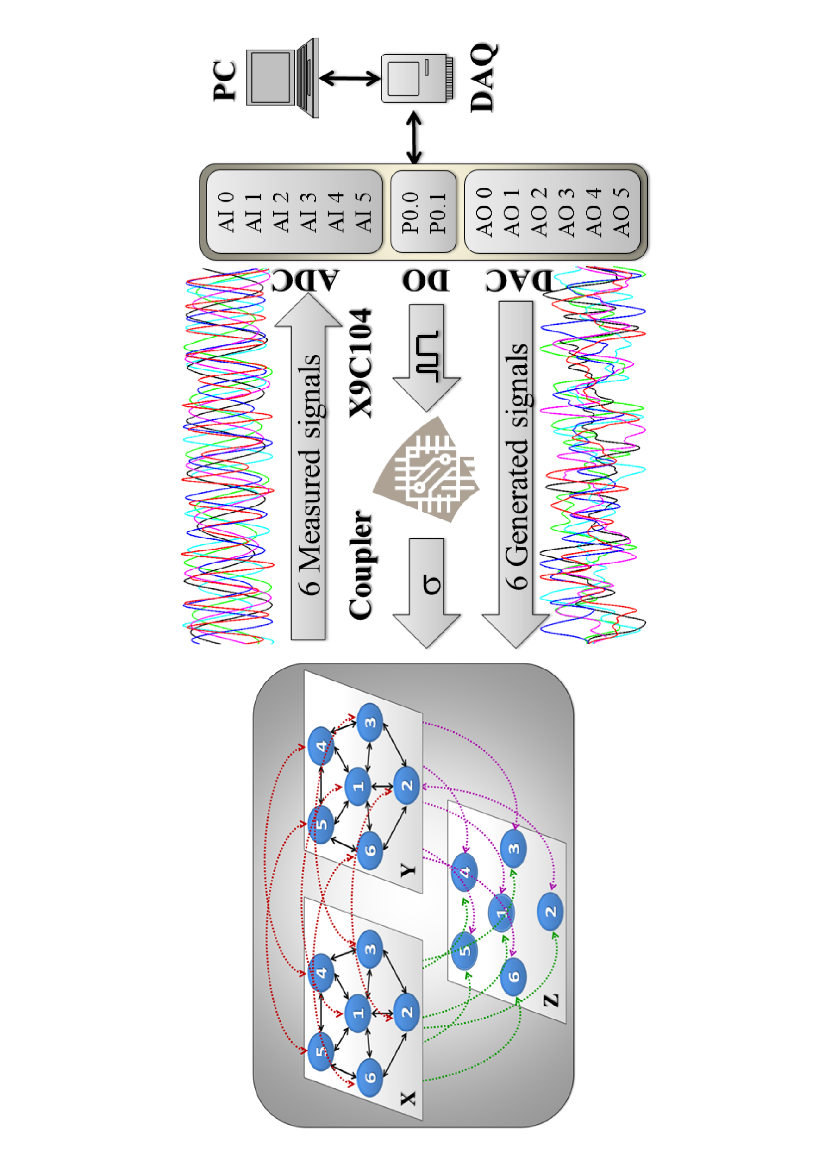}
\vspace{-2cm}
\caption{ \label{fig04} (Color
 online) Experimental setup. On the left  we show a schematic representation of the coupling topology of the 6-circuit network.  The coupling is adjusted using one digital potentiometers X9C104, whose parameters Cu/d (Up/Down resistance) and Cstep (increment of the resistance at each step) are controlled by a digital signal coming from a DAQ Card, P0.0-P0. respectively. The outputs of the circuit are sent to a set of voltage followers that act as a buffer and, then, sent to the analog ports ( AI 0 ; AI 1; ... ; AI 5) of the same DAQ Card. The ports DAC generate the 6 noisy signals to test the robustness of the network. The whole experiment is controlled from a PC with Labview Software.}
\end{figure*}

\section{Numerical simulations}

With the aim of testing the predictions of the MSF, we simulate a network of $N=6$ bidirectionally coupled R\"ossler-like systems, see Fig. \ref{fig01} and Eq. (\ref{system}). We introduce an additive Gaussian noise term (with zero mean) in the $x$ variable whose strength is given by the parameter $\eta$. Next, we calculate the synchronization error $\epsilon$ of the network as a function of the coupling $\sigma$ and  the noise strength $\eta$, for different values of $d_r$. This way, we are able to check the robustness of
the stability of the synchronized manifold depending on how the coupling is distributed among the $x$ and $y$ variables of the system.
To facilitate the comparison with the MSF predictions, we plot the results as a function of $\nu=\sigma \gamma_2$, instead of $\sigma$, being $\gamma_2=2.382$ for the particular network structure shown in Fig. \ref{fig01}. This choice can be justified on the grounds that all the important information from the point of view of the synchronization is given by the MSF value of the eigenmode associated to $\gamma_2$, as the eigenmode associated to $\gamma_6$ is never pushed beyond the boundaries of the synchronization region within the experimentally accessible range of coupling strengths. Figure \ref{fig03} shows the two-dimensional plots $\epsilon(\nu,\eta)$ obtained for $d_r=0$ (coupling only through the $x$ variable), $d_r=0.5$ ($x$ and $y$ couplings are equally active) and $d_r=1$ (coupling only through the $y$ variable). For each of these three cases, the coupling strength $\sigma$ is modified and the synchronization error between all oscillators is calculated (see caption of Fig.\ref{fig03} for details). We consider that the network is out of synchrony for synchronization errors  $\epsilon >0.6$ (red regions in Fig. \ref{fig03}).

When the coupling is introduced through the $x$ variable (Fig. \ref{fig03}.(a)), the R\"ossler-like oscillators behave  as a class III for any value of the noise strength (within the range $0<\eta<5$), i.e. the system only synchronizes for intermediate values of the coupling strength $\nu$. When the noise strength $\eta$ is increased, the synchronization error increases, leading to a complete loss of the synchronization for large values of $\eta$.
For couplings such that $\nu \approx 1.2$ or larger, the network becomes unstable in the sense that the strong coupling makes the individual oscillators abandon the basin of the attractor, and their dynamics blow up. This type of instability is to be expected, as a similar phenomenon is observed in individual R\"ossler-like oscillators for some initial conditions (see, e.g., Ref. \cite{sevilla2015}). 
The reader should notice that no such problems beset the computation of the MSF, as the maximum Lyapunov exponent transverse to the synchronization manifold is computed from effectively one individual oscillator along a trajectory that follows the attractor dynamics. Thus, there is no contradiction in the fact that the MSF determines the attractor dynamics to be synchronizable, while the actual simulation of a network of 6 oscillators never attains attractor dynamics. Nevertheless,  this is an example of the importance of checking  the practical applications of the MSF predictions, specially for high values of coupling strengths.

When coupling is only introduced through the $y $ variable (Fig. \ref{fig03}.(b)) the oscillators behave as class II systems for low values of $\eta$, as expected. Nevertheless, for moderate values of the noise strength ($3 < \eta < 5$) the system shifts to a behavior similar to that of class III systems, synchronizing only for intermediate values of the coupling strength. Finally, when the noise is further increased ($\eta > 5$, not shown here), the network is not able to reach the synchronized state for any value of the coupling, behaving as a class I system. This way, despite being a class II system (when coupled through the y variable), the addition of noise can make system behave differently.

Finally, it is also worth analyzing how the combination of both layers increases the stability of the synchronized manifold. Figure \ref{fig03}.(c) shows the synchronization error for $d_r=0.5$, {\it i.e.} when the coupling is equally distributed among both X and Y layers. We can observe that, as suggested by the theoretical predictions shown in Fig. \ref{fig02}, the combined coupling of the x and y variables enhances the network stability, so that the synchronization of the system is maintained even for high values of the noise strength. Obviously, only for small values of the coupling strength, here measured as $\nu$, unsynchronized motion is observed, as it is expected in class II systems.

Note that the boundaries of the region where synchronization becomes robust show an excellent agreement with the zero crossings of the MSF. On the other hand, in the areas where synchronization is most robust against the presence of noise, the optimal $\nu$ from that point of view is not always close to the minimum of the MSF.
This is not surprising, considering that we are using noise strengths of the order $\eta = 5$, which are certainly beyond any reasonable definition of {\it infinitesimal perturbations}, a necessary requirement for obtaining the MSF. The addition of finite perturbations, being a largely unexplored issue, are beyond of the scope of the MSF framework, and require specific numerical studies adapted to each particular topology, even if some general behaviour has been identified \cite{gutierrez2011}.

\section{Experimental analysis}

\begin{figure}[t!]
\begin{center}
\includegraphics[width=0.45\textwidth]{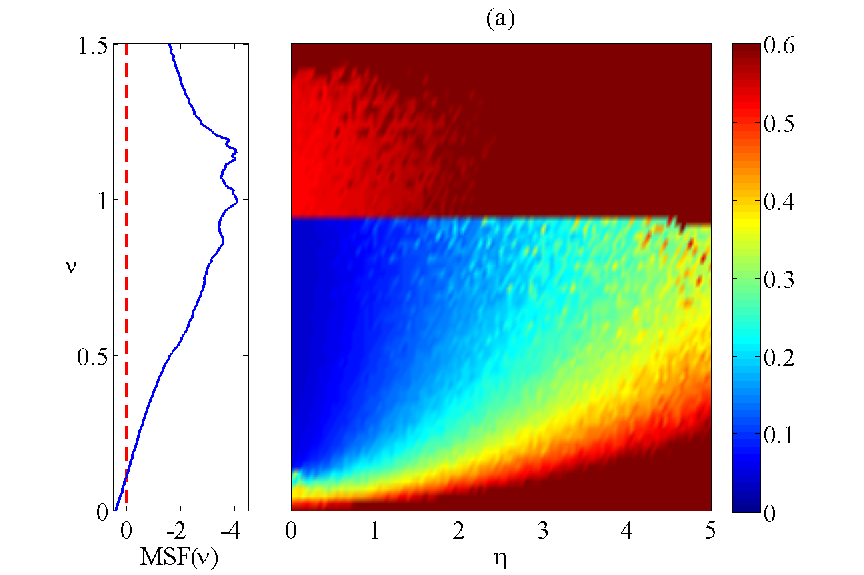}
\includegraphics[width=0.45\textwidth]{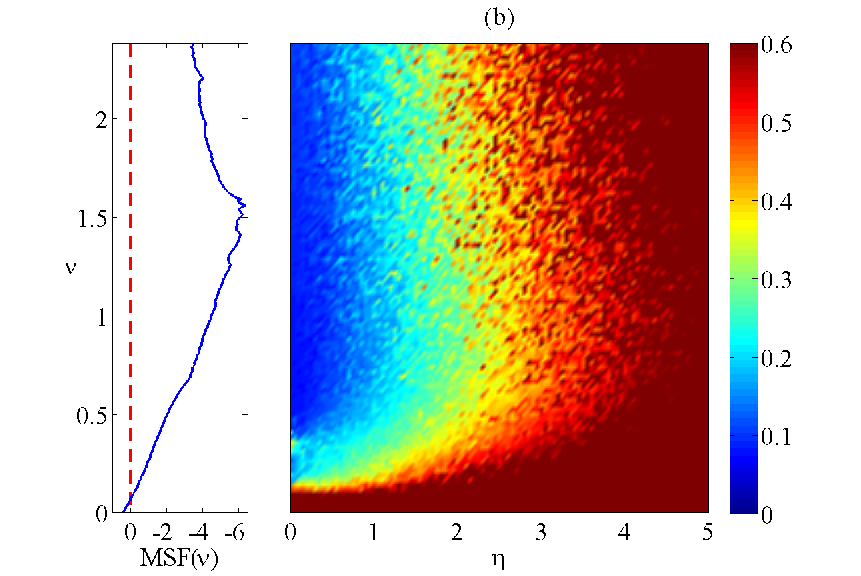}
\includegraphics[width=0.45\textwidth]{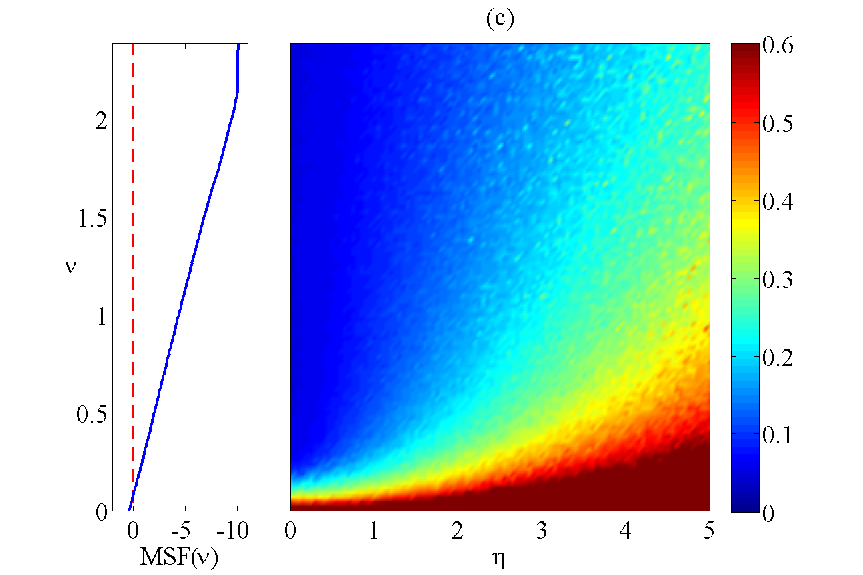}
\end{center}
\caption{\label{fig05}  (Color
 online) Experimental results. 
Synchronization errors as a function of the noise strength $\eta$ and the coupling (normalized to $\nu=d_r  \gamma_2$). As in Fig. \ref{fig03}, panels correspond to: (a) $d_r=0$,  (b) $d_r=1$ and (c) $d_r=0.5$. Region colored in red indicates where the combination of noise and coupling strength leads to a loss of the complete synchronization. The network combining the coupling through the two variables $x$ and $y$ ($d_r=0.5$) is the one showing better performance against noise perturbations. Left plots correspond to the MSF obtained for each value of $d_r$.}
\end{figure}

To test the robustness of the theoretical predictions given by the MSF and the numerical simulations, we have implemented the electronic version of the R\"ossler system described in Eq. \ref{system}. A schematic representation of the experimental design is shown in Fig. \ref{fig04}. It consists of an electronic array (EA), a personal computer (PC) and a multifunction data card (DAQ) composed by 12 analog to digital converters (ADCs) and 6 digital to analog converter (DACs). The ADCs are used for sampling the variable $x$ of the oscillators, while the DACs generate six noisy signals that perturb the dynamics of each node separately. The EA comprises 6 R\"ossler-like electronic circuits forming a spiderweb network (blue nodes), with one central node and five peripheral nodes. Each node has two individual electronic couplers, one for the $x$ variable and the second for the $y$ variable, both controlled by a two digital potentiometers (XDCP), which are adjusted by a digital signal coming from ports P0.0 - 1 (DO). P0.0 is used to increase or decrease the resistance of the voltage divisor ($\sigma$), and P0.1 sets the value of the resistance (the resolution allowing for 100 discretized steps).

The noisy signals are designed in Labview, using the library Gaussian White Noise VI \cite{GWN} that generates six different Gaussian-distributed pseudorandom sequences bounded between $[-1, 1]$. These signals are digitally filtered by a third-order lowpass Butterworth filter \cite{LPF} with a cutoff frequency of $10$ kHz. All the experimental process is controlled from a virtual interface developed in Labview 8.5.

The experimental procedure is the following: first, $\sigma$ and $\eta$ are set to zero and then we introduce the six noisy signals and apply the factor gain ($\eta$). After a waiting time of 500 ms (roughly corresponding to $P=600$ cycles of the autonomous systems), the signals corresponding to the variables of the 6 circuits are acquired by the analog ports (AI 0; AI 1; ... ; AI 5) and the synchronization error is calculated and stored in the PC. Noisy signals are injected by the digital converters (AO 0; AO 1; ... ; AO 5) and this part of the process is repeated 100 times (until the maximum value of $\sigma$ is reached). Finally, $\eta$ is increased to the next value and $\sigma$ is swept again. The whole process is repeated 100 times until the maximum value of $\eta$ is reached.

Figure \ref{fig05} shows the experimental results for a configuration identical to that of the numerical simulations shown in Fig. \ref{fig03}.
We observe that the qualitative agreement between numerics and experiment is excellent, in spite of unavoidable parameter mismatches in
the experimental realization due to the tolerance of the electronic components (between 5$\%$ and 10$\%$). The parameter mismatch, together with the experimental noise, make the oscillators in the network not only slightly different from their mathematical definition, but also non-identical to one another. This way, we confirm experimentally the feasibility of using the MSF for evaluating how the coupling through multiple variables  enhances the stability of the synchronous state of a network under realistic conditions.

\section{Conclusions}


We have seen how an adequate distribution of the coupling strength between the variables of a dynamical system leads to an enhancement of the stability of the synchronized manifold. 
In particular, we have shown that it is possible obtain a MSF that depends on the parameter $d_r$ accounting for the distribution of strength, while maintaining the global coupling constant.  Interestingly, we report the existence of an optimal value of $d_r$ indicating what is the most adequate amount of coupling to be considered at each coupling variable. The optimal value of $d_r$ is independent of the topology of the network, as long as we use the same coupling structure among all variables. 

Using electronic circuits, we have also checked the robustness of the results when noise and parameter mismatch are considered, which confirms the theoretical predictions given by the parametric MSF and, in its turn, reveal that the requirement of the oscillators to be identical can be relaxed.

The proposed framework of decomposing the different dimensions of the system (variables) in interconnected layers paves the way to the use of the multilayer networks tools \cite{kivela2014,boccaletti2014} to further analyze synchronization phenomena in multivariable coupled systems. Indeed, the current theoretical efforts in network theory to define and study complex structures resulting from the interaction of networks, {\it e.g.} interdependent networks and multiplex networks among others, have made great progress in recent times, in showing new emergent phenomena with no counterpart in single (monolayer) complex networks
\cite{aguirre2014,gao2012,aguirre2013,radichi2013}. New developments in that direction could be further extend the results we here present, either using, e.g., the insight developed within the hypernetwork formalism \cite{sorrentino2012,irving2012}, or using new approaches for analyzing
multilayer networks \cite{dedomenico2013}.

Indeed, our methodology has some limitations that must be further explored in the future. First of all, the fact that the coupling must have the same topological structure at all variables is a strict constraint, since real systems may have different configurations depending on the coupling variable. More general, fully multilayer topologies could be considered by resorting to the hypernetwork formalism introduced in \cite{sorrentino2012}, and for greater generality one can use the method in Ref. \cite{irving2012}. With this methodology, the case of 
$\mathcal{A}^X \neq \mathcal{A}^Y$ could be addressed at the cost of introducing some more complexity to the problem. Nevertheless, it would
be of great interest, since it would raise new questions such as what the adequate combination of topologies would be given a specific distribution of weights $d_r$. Second, since the parametric MSF depends on the dynamical system implemented in the network, we can not guarantee the existence of an optimal balance of the distribution of coupling between layers in other dynamical systems, at least until their corresponding MSF have been analyzed.

\section{Acknowledgements}

Authors acknowledge D. Papo and P.L. del Barrio for fruitful conversations. Support from MINECO
through projects FIS2011-25167, FIS2012-38266 and FIS2013-41057-P is also acknowledged. AA and JGG acknowledge support from the EC FET-Proactive Project PLEXMATH (grant 317614) and MULTIPLEX (grant 317532). JGG acknowledges support from MINECO through the Ram\'on y Cajal program, the Comunidad de Arag\'on (Grupo FENOL) and the Brazilian CNPq through the PVE project of the Ciencia Sem Fronteiras program. AA acknowledges ICREA Academia and the James S.\ McDonnell Foundation. R.S.E. acknowledges Universidad de Guadalajara, CULagos (Mexico) for financial support (OP/PIFI-2013-14MSU0010Z-17-04, PROINPEP-RG/005/2014, UDG-CONACyT/I010/163/2014) and CONACyT (Becas Mixtas MZO2015/290842).


\begin{thebibliography}{}

\bibitem{boccaletti2006} S. Boccaletti, V. Latora, Y. Moreno, M. Chavez and D.-U. Hwang,
{\it Phys. Rep.} {\bf 424},175-308 (2006).

\bibitem{arenas2008} A. Arenas, A. D\'{\i}az-Guilera, J. Kurths, Y. Moreno and C. Zhou,
{\it Phys. Rep.} {\bf 469}, 93-153 (2008).

\bibitem{barrat2008} A. Barrat, M. Barthelemy and A. Vespignani. {\it Dynamical Processes on Complex Networks}, Cambridge University Press (2008).

\bibitem{pecora1998} L.M. Pecora and T.L. Carroll,
{\it Phys. Rev. Lett.} {\bf 80}, 2109 (1998).

\bibitem{motter2005a} A.E. Motter, C. Zhou and J. Kurths,
{\it Phys. Rev. E} {\bf 71}, 016116 (2005).

\bibitem{chavez2005} M. Chavez, D.-U. Hwang, A. Amann, H. G. Hentschel and S. Boccaletti,
{\it Phys. Rev. Lett.} {\bf 94}, 218701 (2005).

\bibitem{motter2005b} A. E. Motter, C.S. Zhou and J. Kurths,
{\it Europhys. Lett.} {\bf 69}, 334-340 (2005).

\bibitem{nishikawa2005} T. Nishikawa and A.E. Motter,
{\it Phys. Rev. E} {\bf 73}, 065106 (2006).

\bibitem{donetti2005} L. Donetti, P. I. Hurtado and M. A. Mun\~oz,
{\it Phys. Rev. Lett.} {\bf 95}, 188701 (2005).

\bibitem{sorrentino2012} F. Sorrentino,
{\it New J. Phys.}  {\bf 14}, 033035 (2012).

\bibitem{laplacian}  A network with a connectivity matrix $\mathcal{A} = \{a_{ij}\}$ ($a_{ij} =1$ if nodes $i$ and $j$ are connected, and 0 otherwise) can be represented by the Laplacian matrix $\mathcal{L}=\{l_{ij}\}$, where $l_{ij} = \delta_{ij} (\sum_{k=1}^N  a_{ik}) - a_{ij}$ ($\delta_{ij}$ denoting the Kronecker delta).

\bibitem{irving2012} D. Irving and F. Sorrentino,
{\it Phys. Rev. E} {\bf 86}, 056102 (2012).

\bibitem{modelo} A.N. Pisarchik, R. Jaimes-Re\'ategui, J.R. Villalobos-Salazar, J.H. Garc\'ia-L\'opez and S. Boccaletti,
{\it Phys. Rev. Lett.} {\bf 96}, 244102 (2006).

\bibitem{bragard2003} J. Bragard, S. Boccaletti and H. Mancini,
{\it Phys. Rev. Lett.} {\bf 91}, 064103 (2003).

\bibitem{hwang2005} D.-U. Hwang, M. Chavez, A. Amann and S. Boccaletti,
{\it Phys. Rev. Lett.} {\bf 94}, 138701 (2005).

\bibitem{huang2009} L. Huang, Q. Chen, Y.-C. Lai and L.M. Pecora,
{\it Phys. Rev. E} {\bf 80}, 036204 (2009).

\bibitem{aguirre2014}  J. Aguirre, R. Sevilla-Escoboza, R. Guti\'errez, D. Papo and J. M. Buld\'u,
{\it Phys. Rev. Lett.} {\bf 112}, 248701 (2014).

\bibitem{sevilla2015} R. Sevilla-Escoboza, J. M. Buld\'u, A. N. Pisarchik, S. Boccaletti, and R. Guti\'errez,
{\it Phys. Rev. E} {\bf 91}, 032902 (2015).

\bibitem{gutierrez2011}
R. Guti\'errez, F. del-Pozo and S. Boccaletti,
{\it PLoS ONE} \textbf{6}, e20236 (2011).

\bibitem{GWN} National Instrument, {\bf Gaussian White Noise VI} \url{http://zone.ni.com/reference/en-XX/help/371361J-01/lvanls/gaussian_white_noise/}

\bibitem{LPF} National Instrument, {\bf Filter} \url{http://zone.ni.com/reference/en-XX/help/371361J-01/lvexpress/signal_filter/}

\bibitem{kivela2014} M. Kivel\"a, A. Arenas, M. Barthelemy, J.P. Gleeson, Y. Moreno, M.A. Porter,
Multilayer Networks
J. Complex Netw. {\bf 2}, 203 (2014)

\bibitem{boccaletti2014} S. Boccaletti, G. Bianconi, R. Criado, C.I. del Genio, J. G\'omez-Garde\~nes, M. Romance, I. Sendi\~na-Nadal,
Z. Wangk, M. Zanin, Phys. Rep. {\bf 544}, 1 (2014).

\bibitem{gao2012}  J. Gao, S.V. Buldyrev, H.E. Stanley and S. Havlin,
{\it Nat. Phys.} {\bf 8}, 40 (2012).

\bibitem{aguirre2013} J. Aguirre, D. Papo and J. M. Buld\'u,
{\it Nat. Phys.} {\bf 9}, 230 (2013).

\bibitem{radichi2013}  F. Radichi and A. Arenas,
{\it Nat. Phys.},  {\bf 9} 717 (2013).

\bibitem{dedomenico2013}
M. De Domenico,  A. Sol\'e-ribalta, E. Cozzo, M. Kivel\"a, Y. Moreno, M.A. Porter, S. G\'omez and A. Arenas, Phys. Rev. X {\bf 3}, 041022 (2013).

\end{thebibliography}
\end{document}